\documentclass[journal=nalefd,manuscript=letter]{achemso}

\usepackage[version=3]{mhchem} 
\usepackage{graphicx}

\usepackage[colorinlistoftodos]{todonotes}

\author{Leopold Talirz}
\author{Hajo S\"ode}
\affiliation{nanotech@surfaces laboratory, Empa, Swiss Federal Laboratories for Materials Science and Technology, 8600 D\"ubendorf, Switzerland}
\author{Shigeki Kawai}
\affiliation{International Center for Materials Nanoarchitectonics, National Institute for Materials Science, 1-1, Namiki, Tsukuba, Ibaraki 305-0044, Japan.}
\author{Pascal Ruffieux}
\affiliation{nanotech@surfaces laboratory, Empa, Swiss Federal Laboratories for Materials Science and Technology, 8600 D\"ubendorf, Switzerland}
\author{Ernst Meyer}
\affiliation{Department of Physics, University of Basel, Klingelbergstrasse 82, CH-4056 Basel, Switzerland.}
\author{Xinliang Feng}
\author{Klaus M\"ullen}
\affiliation{Max Planck Institute of Polymer Research, Mainz, Germany}
\author{Roman Fasel}
\affiliation{nanotech@surfaces laboratory, Empa, Swiss Federal Laboratories for Materials Science and Technology, 8600 D\"ubendorf, Switzerland}
\affiliation{Department of Chemistry and Biochemistry, University of Bern, 3012 Bern, Switzerland}
\author{Carlo A. Pignedoli}
\affiliation{nanotech@surfaces laboratory, Empa, Swiss Federal Laboratories for Materials Science and Technology, 8600 D\"ubendorf, Switzerland}
\email{carlo.pignedoli@empa.ch}
\author{Daniele Passerone}
\affiliation{nanotech@surfaces laboratory, Empa, Swiss Federal Laboratories for Materials Science and Technology, 8600 D\"ubendorf, Switzerland}
\email{daniele.passerone@empa.ch}

\title[From Molecules to Ribbons: Length and Terminus Dependence of the Electronic Band Gap]
  {Band gap of atomically precise graphene nanoribbons as a function of ribbon length and termination}

\abbreviations{DFT,STM, STS}

\begin{document}


\begin{abstract}
We study the band gap of finite $N_A=7$ armchair graphene nanoribbons (7-AGNRs) on Au(111) 
through scanning tunneling microscopy/spectroscopy combined with density functional theory calculations. 
The band gap of 7-AGNRs with lengths of 6 nm and more is converged to within 0.1 eV of its bulk value of 2.3 eV, while 
the band gap opens by several hundred meV in very short 7-AGNRs.
The termination has a significant effect on the band gap, 
doubly hydrogenated termini yielding a lower band gap than singly hydrogenated ones. 

\end{abstract}


One of the driving motivations for investigating the electronic properties 
of two-dimensional materials, such as graphene, lies in their beneficial electrostatics for field effect transistor applications, 
which should enable downscaling of the channel length below conventional limits\cite{Schwierz2010}.
While large-scale graphene is a semimetal, graphene nanoribbons (GNRs) 
exhibit an electronic band gap that depends both on the width of the ribbon 
and on the atomic structure of the ribbon edge. 

While many studies, both theoretical and experimental, have addressed the 'bulk' electronic structure as well and optical properties of GNRs \cite{son_energy_2006,prezzi_optical_2008,yang_two-dimensional_2008,yazyev_emergence_2010,wakabayashi_electronic_2010,zhu_scaling_2011,kimouche_ultra-narrow_2015,kharche_width_2016} (see also \cite{andreoni_electronic_2018} and references therein) in view of future applications in electronic devices at the nanoscale, GNRs of finite length are of considerable interest as well. 
A recently developed bottom-up approach, based on surface-assisted colligation and subsequent 
dehydrogenation of specifically designed precursor monomers, 
enables the synthesis of some prototypical GNRs with atomic precision\cite{Cai2010}.
Here, we investigate the evolution of their electronic structure as a function of the number of monomers that make up the ribbon. Since the GNRs are free of defects, 
the findings reflect intrinsic properties of the class of GNRs under study and enable direct comparison with ab initio calculations. In a second step, we study the electronic effect of introducing a specific atomic defect at the terminus. 

\begin{figure}
  \includegraphics[width=0.75\textwidth]{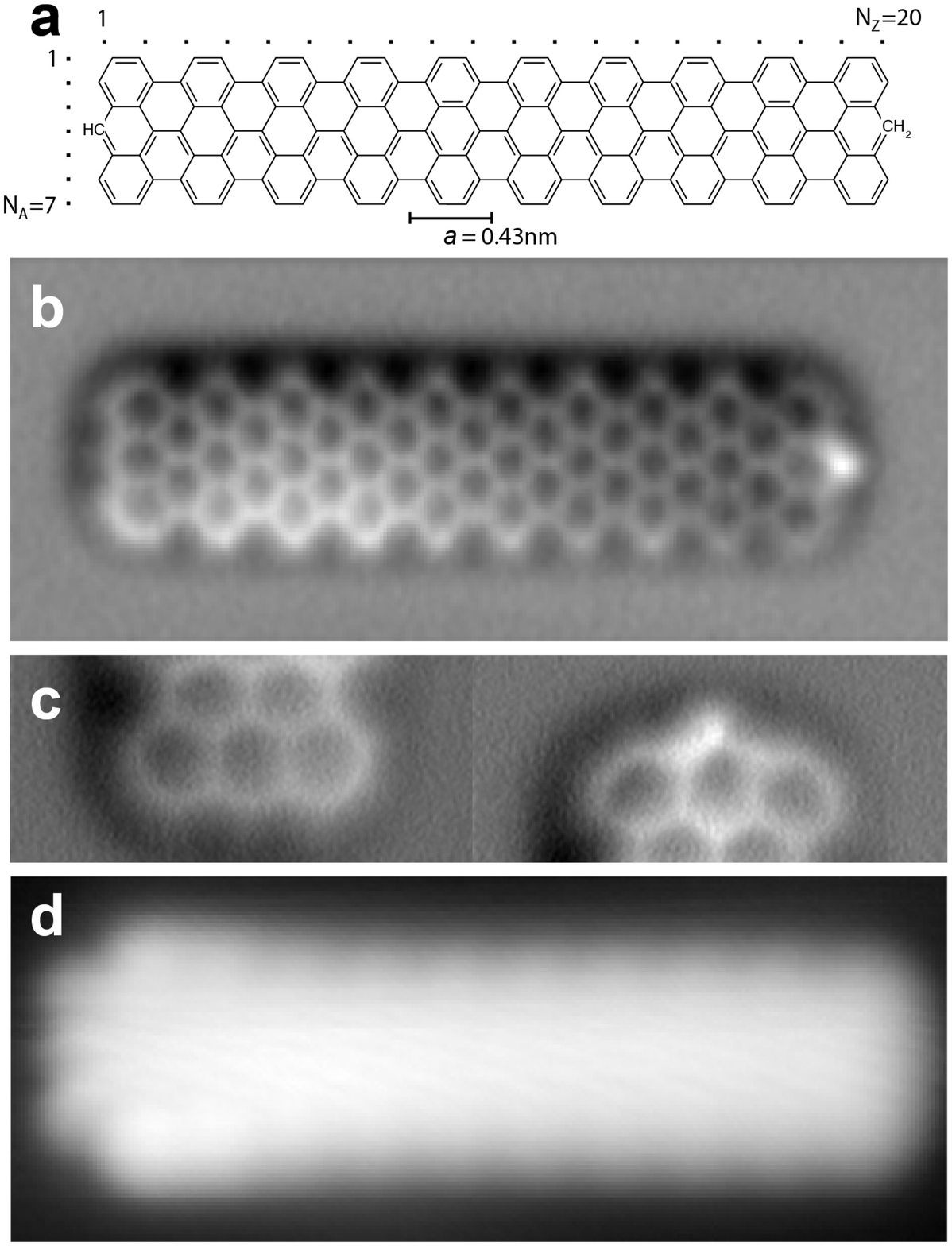}
  \caption{Atomic structure of 7-AGNR. a) Scheme of 7-AGNR with length $L=10a \approx 4.3\,\textrm{nm}$. b) Non-contact AFM image. Oscillation amplitude A =  43 pm, bias voltage V= 0 mV . c) Details of the termini structures from AFM; d) STM image acquired with a tunneling current setpoint I=10 pA and tip bias V$_{\mathrm{tip}}$=-200 meV.}
  
  \label{fig:structure}
\end{figure}

Figure \ref{fig:structure} shows a $N_A=7$ armchair graphene nanoribbon (7-AGNR) imaged by non-contact atomic force microscopy (AFM) as well as scanning tunneling microscopy (STM).
While the long edges of the 7-AGNR are monohydrogenated, the GNRs may be terminated by one hydrogen at the central carbon atom (\ce{CH}, left terminus) or by two hydrogen atoms (\ce{CH2}, right terminus), depending on the temperature $T_c$ chosen for the dehydrogenation step~\cite{Blankenburg2012,Talirz2013,Lit2013}.
We thus consider the additional hydrogen an atomic defect.
By varying $T_c$ from $300^\circ$ to $400^\circ$, the relative probability of \ce{CH} and \ce{CH2}-termination can be tuned from a majority of \ce{CH2} to almost exclusively \ce{CH}. Unless otherwise stated, in the following we consider GNRs that have the same termination at both ends. Both termination and length of the GNRs are easily identified not only in AFM, but also in STM\cite{Talirz2013,Lit2013}, which is the instrument used in the rest of the study.
The complete knowledge of the GNRs' \emph{atomic} structure forms the basis for the discussion of their \emph{electronic} structure.

One fundamental question we would like to address here is: how does the transition from a molecule to a nanoribbon occur and at which length can the electronic structure be considered \emph{converged}?
This question relates not only to possible applications of GNRs in nanoscale devices. It also concerns scientific studies using spectroscopic techniques such as (inverse) photoemission, which measure the average electronic structure of many GNRs, and where effects due to finite length of the GNRs generally are to be avoided.
Naturally, the answer to this question depends on the quantity of interest and in this study we restrict ourselves to the electronic band gap.

We have measured the band gap of finite 7-AGNRs on Au(111) using scanning tunneling spectroscopy (STS).
The STM tip was positioned in the center between the two termini and differential conductance spectra were recorded on a line scan across GNR using a lock-in amplifier, which modulated the voltage at $860\,\textrm{Hz}$ with $16\,\textrm{mV}$ root-mean-square amplitude. The spectra were recorded by ramping the voltage from -1.5 V to 2.0 V at constant tip-sample distance set by $I=0.2$ nA and $V = -1.5$ V. Following reference \cite{Ruffieux2012}, the band gap was extracted using the half-maximum of the valence and conduction band onsets.

\begin{figure}
  \includegraphics[width=0.5\textwidth]{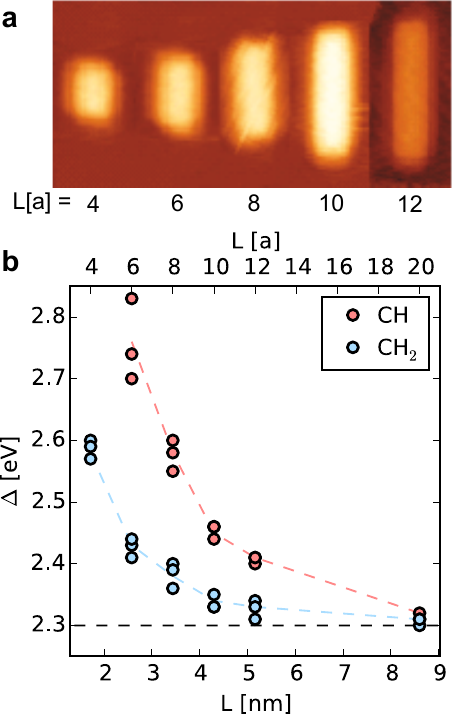}
  \caption{Band gap of 7-AGNRs. a) STM topography of finite 7-AGNRs (V$_{{\mathrm tip}}$ = 1.5 V, I = 0.1 nA). b) Band gap of 7-AGNRs as determined experimentally (markers). The dashed
lines follow the averaged values for each length. The black dashed line at 2.3 eV marks the bulk value.}
  \label{fig:bg-vs-length}
\end{figure}

As shown in Figure \ref{fig:bg-vs-length}, the band gap increases with decreasing length of the GNR.
For GNRs longer than $8\,\textrm{nm}$, the band gap is converged to within $50\,\textrm{meV}$.
And in the length range investigated here, we note two different trends that we attribute to the different termination of the ribbons: we claim that the \ce{CH} terminated
ribbons show a significantly higher band gap than \ce{CH2}-terminated ones of the same length. We demonstrate this claim in the following.

\begin{figure}
  \includegraphics[width=1.0\textwidth]{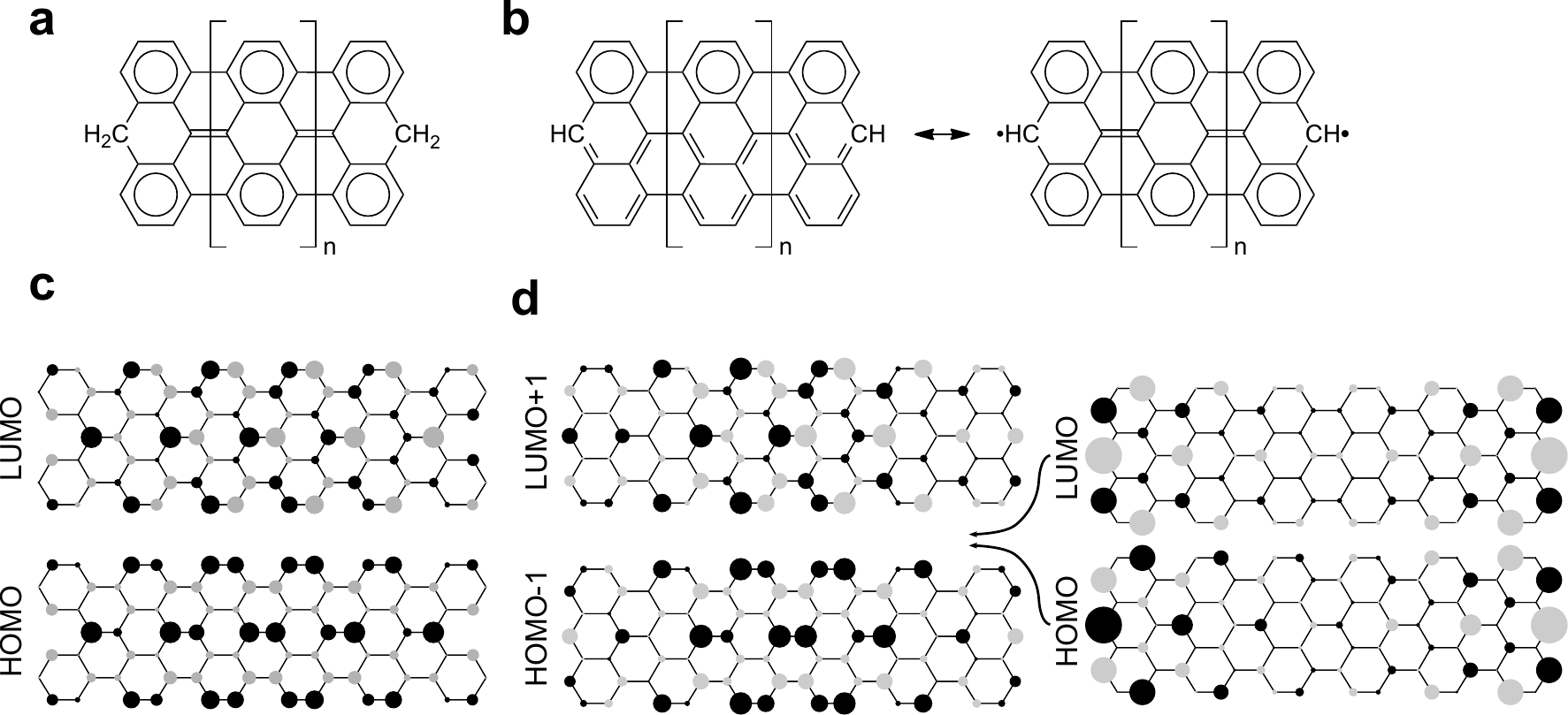}
  \caption{Electronic structure of 7-AGNR. a) Clar formula for \ce{CH2} termination. b) Clar formulas for \ce{CH} termination. c) Tight binding orbitals of \ce{CH2}-terminated 7-AGNR with length $L=6a$. The circle area is proportional to the electron density, gray/black indicates the sign of the wave function. d) Same as c), but for \ce{CH} termination.}
  \label{fig:clar-tb}
\end{figure}

In order to rationalize this observation, it is helpful to start with some basic insights into the electronic structure of graphene nanoribbons provided by Clar's theory of the aromatic sextet\cite{Clar1972,Wassmann2010}. 
Clar's rule states that the representation of a hydrocarbon with the highest number of aromatic sextets correspond to the lowest energy.
Figure \ref{fig:clar-tb} a) shows the Clar formula for the \ce{CH2}-terminated 7-AGNR. The Clar formula is unique and it contains two aromatic sextets per unit cell of the GNR\cite{Talirz2013,Konishi2013b}.
The case of CH-terminated 7-AGNRs is more complicated. Multiple Clar formulae exist and they have only one aromatic sextet per repeat unit, leading to considerably reduction in aromatic stabilization compared with the \ce{CH2} case. However, at the cost of introducing one unpaired electron near each terminus of the GNR, a unique Clar formula with two Clar sextets per repeat unit are achieved also in this case.
The electronic structure for the \ce{CH} termination is thus characterized by a competition between the cost of breaking a $\pi$-bond, favoring structure 1, and the energetic stabilization due to increased aromaticity, favoring structure 2. Since the latter increases linearly with GNR length, the question arises at which length structure 2 will dominate.
Konishi et al. have studied CH-terminated 7-AGNRs of lengths $L=2a,3a$ and $4a$ using the complete active space self-consistent field (CASSCF) method and found a biradical character of $7, 54$ and $91\%$, this indicating that the transition occurs between $L=2a$ and $L=4a$ \cite{Konishi2013}.
Similarly, an estimate based on the experiments in thermochemistry used in \cite{yeh_role_2016} as a phenomenological guide to assess the stability of aromatic vs. radical structures, namely a gain of 90 kJ/mol per aromatic ring vs. a penalty of 272 kJ/mol per couple of unpaired electrons, gives a threshold of $L=3a$.
For the length of 7-AGNRs considered here, starting from $L=4a$, the Clar structure is therefore dominant. 
We note that this is also in agreement with spin-unrestricted DFT calculations using the PBE approximation to the exchange functional, which find a spin-unpolarized ground state for $L=2a$, while spin-polarization appears for $L=3a$ and above.

While this explains the edge-localized states associated with the \ce{CH} termination, it does not yet explain the influence of the termination on the band gap.
At this point, a proper definition of the term 'band gap' is required. In solid state physics, the band gap of an infinite crystal is defined as the energy difference between the top of the valence band and the bottom of the conduction band. For a finite crystal, the energy bands become sets of discrete energy levels and we define the band gap as the energy difference between the highest such energy emerging from the valence band and the lowest such energy emerging from the conduction band. While the corresponding states are delocalized over essentially the whole crystal, some surfaces give rise to in-gap states localized near the surface. These surface states are disregarded when computing the band gap. In the language of molecular chemistry, the energy difference between the lowest unoccupied molecular orbital (LUMO) and the highest occupied molecular orbital (HOMO) therefore equals the band gap only in the absence of surface states.
Graphene nanoribbons are one-dimensional crystals and in the case of the 7-AGNR, the \ce{CH} termination is associated with a Tamm state, which is saturated by addition of another hydrogen in the \ce{CH2} case. The bandgap of the \ce{CH}-terminated 7-AGNRs is therefore given by the energy difference between LUMO+1 and HOMO-1, while it coincides with the HOMO-LUMO gap for the \ce{CH2}-termination. We note that this definition of the bandgap is suitable in particular in view of electronic transport applications of GNRs, where states that are localized near the termini of the GNRs do not contribute. It is also the gap measured when positioning the STM tip in the center between the two termini.

Martinez et al. \cite{Martin-Martinez2012a} argue that the aromatic structure is associated with a smaller band gap than the K\'{e}kule structure due to stronger electron delocalization. Even if this is true, it is not at all clear that this is the reason behind the difference in band gap.
According to Konishi et al. \cite{Konishi2010}, the diradical character of 4-anthryl is already $91\%$, while the difference in band gap to be explained is of the order of $25\%$.
While the trend from the Clar structures goes into the right direction, it may actually not be the main mechanism behind the difference in band gap.

\begin{figure}
  \includegraphics[width=0.75\textwidth]{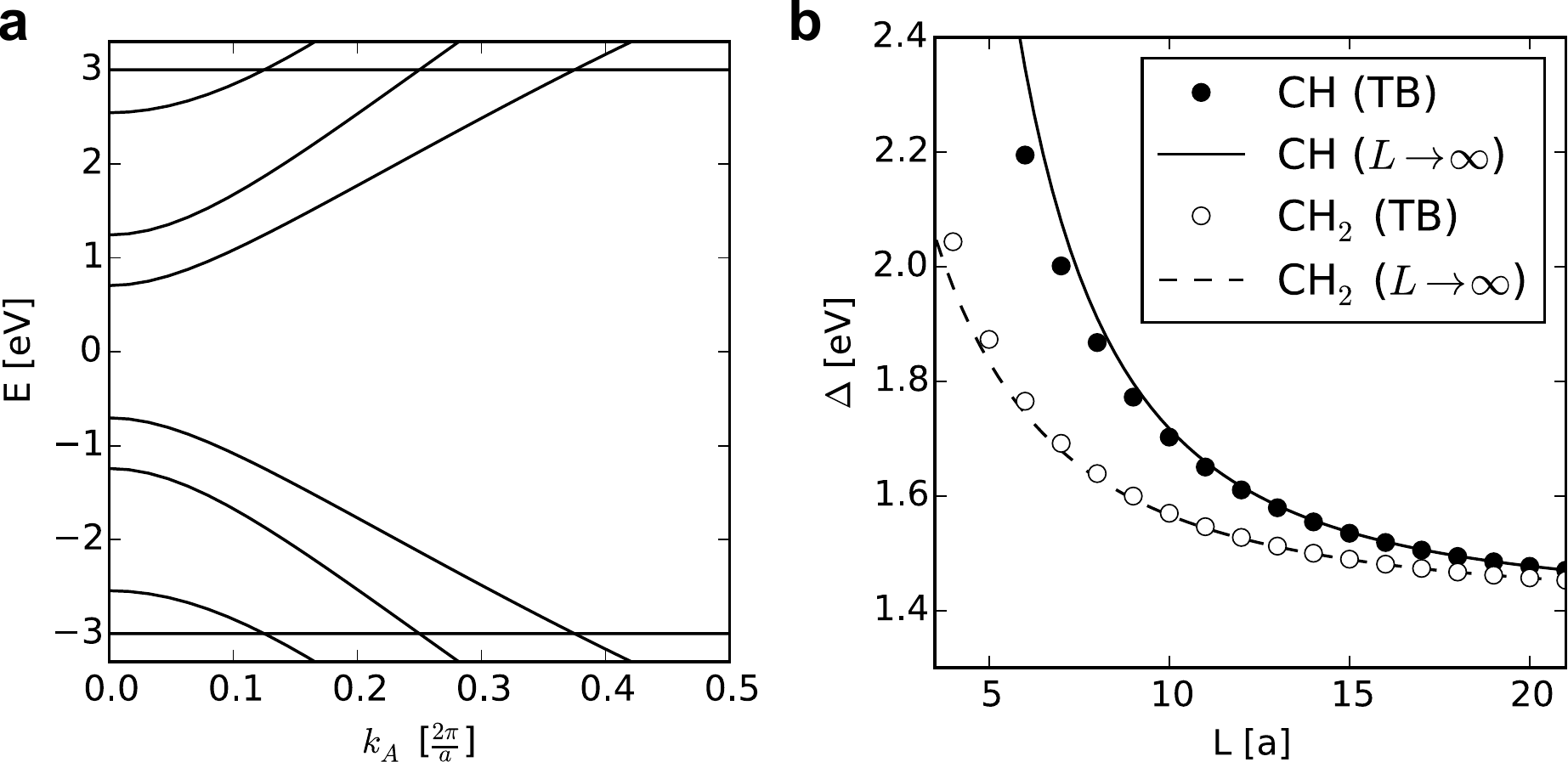}
  \caption{Tight binding band gap. a) Energy bands of 7-AGNR using  hopping integral $t=-3\,\textrm{eV}$. b) Band gap as function of length and termination, at the TB (markers) and DFT (lines) level.}
  \label{fig:tb}
\end{figure}

A simple model that captures the effect on a qualitative level, is tight-binding using one $\pi$-orbital per carbon atom and considering hopping only between nearest neighbors. 
Since the central carbon atom at the terminus does not contribute an electron to the $\pi$-system when forming bonds with two hydrogen atoms, the \ce{CH2}-termination is equivalent to a so-called 'cove defect', where the central carbon atom is simply removed\cite{Wakabayashi2010a}.
Having a closer look at the states for the \ce{CH2} and \ce{CH}-termination in Figure \ref{fig:clar-tb} c),d) it appears that the \ce{CH} states are slightly pushed away from the termini compared to their \ce{CH2} counterparts. Stronger confinement would naturally lead to a larger band gap and in the following we quantify this effect based on the analytic expressions derived by Wakabayashi for GNRs with monohydrogenated zigzag and armchair edges\cite{Wakabayashi2010}.
The delocalized states of the finite system can be related to the dispersion of graphene $E(\vec{k})$, by finding a corresponding wave vector $\vec{k}=(k_Z,k_A)$ with components $k_Z$ along the zigzag and $k_A$ along the armchair direction\cite{Wakabayashi2010}.
While confinement by the armchair edges gives rise to a simple discretization
 \[ k_Z = r_A \frac{\pi}{(N_A+1)a_Z}\quad\textrm{with} \ r_A\in\{1,\ldots,N_A\},\ a_Z=a/\sqrt{3},\]
 confinement along the armchair direction  results in a transcendental equation for $k_A(k_Z)$. 
As derived in \cite{TalirzPhd}, in the limit of small $k_A$ this equation simplifies to
 \[ k_A\approx r_Z\frac{\pi}{(N_Z+\delta)a_A}\quad \textrm{with}\ r_Z\in\{1,\ldots\},\ r_Z\ll N_Z,\ a_A=a/2 \] 
with $\delta=\left(1+1/2\cos[\frac{r_Z}{N_A+1}\pi]\right)^{-1}$.
For the frontier bands of the \ce{CH}-terminated 7-AGNR ($r_Z=5$), we obtain $\delta\approx -3.3$.
For the \ce{CH2}-terminated GNRs, where the analytic expressions from Ref. \cite{Wakabayashi2010} do not apply, fitting of the numerical tight binding values with the same formula yields $\delta\approx +4.0$.
Coming back to the original question, the different terminations can thus be viewed as imposing a different effective 'electronic length', the \ce{CH}-terminated GNRs appearing 
$(4.0+3.3)a_A\approx 1.6\,\textrm{nm}$ shorter than the \ce{CH_2}-terminated ones, while consisting of the same number of monomers.
Figure \ref{fig:tb} b) compares the band gaps calculated with the tight-binding model for a hopping integral $t=-3\,\textrm{eV}$ to the results using the above-mentioned approximation for $k_A$. The prediction of the simple "effective electronic length" approximation agrees remarkably well with the full tight-binding calculation.

While single-orbital nearest-neighbor tight binding lends itself to qualitative investigations, neglecting the variation of on-site energies\cite{Son2006}, hopping between more than nearest neighbors and the Coulomb repulsion between electrons\cite{Ervasti2013} corresponds to a serious simplification.
In this respect, an ab initio scheme such as density functional theory (DFT) presents a significant step forward. It has been demonstrated that the effective single-particle description of spin-unrestriced Kohn-Sham DFT captures the essential features as compared to a many-body description using the Hubbard model\cite{Ervasti2013}.
We have performed spin-unrestricted DFT calculations using the PBE generalized gradient approximation to the exchange-correlation functional\cite{Perdew1996}. 
We have used the CP2K code, expanding the Kohn-Sham wave functions on an atom-centered Gaussian basis set of triple-zeta double polarization (TZV2P) quality
and using a cutoff energy of $350\,\textrm{Ry}$ for the plane-wave representation of the charge-density. 
Atomic positions were relaxed until the forces were below $3\,\textrm{meV}/$\AA\ . The Martyna-Tuckerman Poisson solver\cite{Martyna1999} was used to decouple periodic images.

\begin{figure}
  \includegraphics[width=0.75\textwidth]{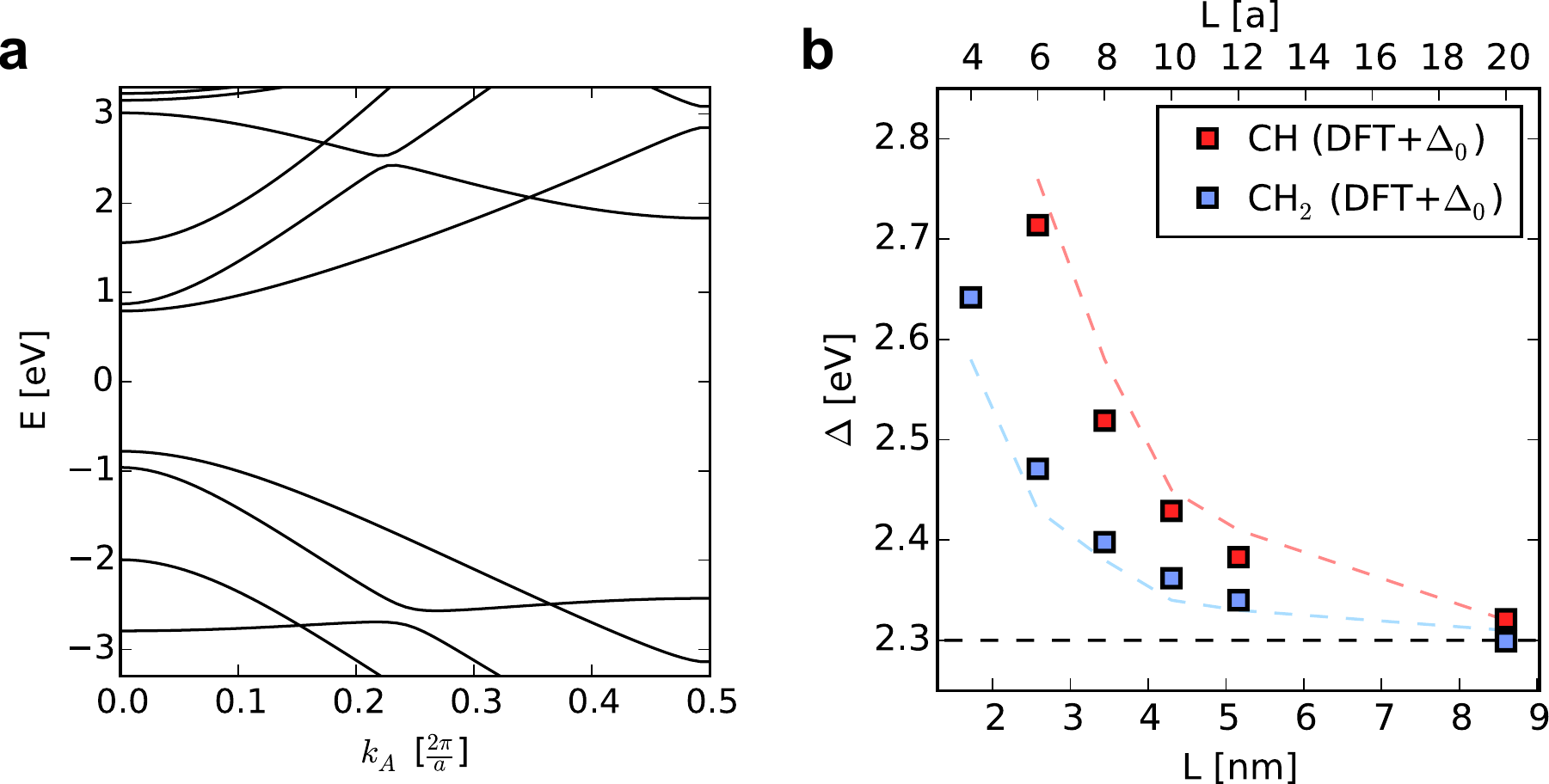}
  \caption{DFT electronic structure. a) Energy bands of 7-AGNR. The center of the gap has been shifted to zero energy.  b) DFT band gap as function of length and termination, increased by a constant $\Delta_0=0.73\,\textrm{eV}$ to aid comparison with experimental results. The dashed lines are the same as in Figure \ref{fig:bg-vs-length}, namely the averages of experimental values and the bulk value of 2.3 eV.}
  \label{fig:dft}
\end{figure}

While Kohn-Sham DFT with standard semi-local exchange-correlation functionals is well-known to understimate band gaps, we find that correcting the DFT band gap, obtained by the band structure in Figure  \ref{fig:structure} a), by a constant shift of
\[ \Delta_0=2.3\,\textrm{eV}-1.57\,\textrm{eV}=0.73\,\textrm{eV}\quad,\]
defined as the band gap difference between experiment and DFT for $L\rightarrow\infty$, yields good agreement with experiment, as shown in Figure \ref{fig:dft} b).
While there may be no general justification for such an approach, the tight-binding analysis demonstrates that a major contribution to the band gap opening is directly related to the band dispersion. Since the band dispersion from Kohn-Sham DFT is in good agreement with the experimental band dispersion of 7-AGNRs on Au(111)\cite{sode_electronic_2015}, it seems reasonable to expect that the opening of the  band gap with decreasing length is well described.


Finally, we mention two caveats that apply to this study.
The first concerns the Tamm states, whose description is relevant here only as far as its influence on the delocalized states is concerned.
DFT predicts a spin-splitting of the Tamm states, which is not observed in STS experiments of 7-AGNRs on Au(111), where a single peak is found at $E-E_F\approx +50\,\textrm{meV}$\cite{VanderLit2013}. 
This finding has been explained by a transfer of electrons from the GNR to the Au substrate, noting that the the splitting of the Tamm states is strongly reduced in DFT calculations of charged 7-AGNRs in vacuum\cite{VanderLit2013}.
While charging of the 7-AGNR with $+2|e|$ (emptying the Tamm states at both ends) would indeed influence the band gap; in this case DFT also predicts the energy of the empty Tamm states to move down in energy, merging with the valence band for $L\rightarrow \infty$.
This is clearly not the case in experiment, where onsets of valence and conduction band are found at $-0.8\,\textrm{eV}$ and $+1.5\,\textrm{eV}$ respectively\cite{sode_electronic_2015}.
This seems to suggest an intermediate scenario of small fractional charge transfer and strong screening from the substrate.
We note that this is problematic to address ab initio, both due to the known problems of local- and semilocal functionals in accurately describing charge transfer as well as due to the difficulty in determining the correct adsorption geometry (intermediate between physisorbed and chemisorbed state at the zigzag edge).

Second, it is possible that the STS signal used to determine the band gap is dominated by contributions from states of the CB+1\cite{sode_electronic_2015}.
Assuming that the relative intensity of the orbitals in STS is given by the LDOS at the tip (Tersoff-Hamann approximation), orbitals arising from the CB appear much fainter than orbitals arising from the CB+1. 
Since the CB and CB+1 lie energetically very close in the range of k-vectors probed here, this gives rise to a possible uncertainty in the exact determination of the band gap.

We also note that in the case of 7-AGNR the convergence with length to the bulk value of the electronic band gap is reached already for short segments. This is not the case for ribbons of different width, like the so-called quasi-metallic GNRs where the width $L=n\times a\,n=3p+2$ with $p$ integer. In that case, as discussed in \cite{wang_quantum_2017} much longer ribbon segment are necessary (beyond 30 $nm$) to converge to the ``bulk'' value of the band gap, due to the smaller effective mass of the charge carriers in the latter case. It would be interesting to see if the ``effective length'' correction would remain meaningful also in these other cases. 

In conclusion, we have proved both experimentally and theoretically the sensitivity of the electronic band gap to the termini structures of finite-size armchair ribbons of different lengths. The experimental observations are rationalized with a simple model that defines an "effective length" of the ribbon based on the $CH_2$ or $CH$ decoration of the GNR termini, namely, the $CH$-terminated ribbons appear $\approx \, 1.6$ nm shorter than the $CH_2$-terminated ones with the same number of monomers. We demonstrated that simple computational schemes can provide valuable insight in the description of trends of the electronic properties of this class of nanomaterials as a function of their size and termination, and  higher level methods need to enter into play only when quantitative values for applications are needed.



\begin{acknowledgement}

Calculations were supported by a grant from the Swiss National Supercomputing Centre (CSCS) under project ID s746. The Swiss National Science Foundation and the Swiss National Centre for Computational Design and Discovery of Novel Materials (MARVEL) are acknowledged for financial support.
\end{acknowledgement}
\section{Methods}

All measurements were performed with a commercially available Omicron low temperature scanning tunneling microscopy (STM)/atomic force microscopy (AFM) system, operating in ultra-high vacuum at 4.8 K. We used a tuning fork with a chemically etched tungsten tip as a force sensor.~\cite{Gie98} The resonance frequency and the mechanical quality factor are 23067.3~Hz and 26704, respectively. The high-stiffness of 1800~N/m realizes a stable operation with small amplitude.~\cite{Gie03} The frequency shift, caused by the tip-sample interaction, was detected with a commercially available digital phase-locked loop (Nanonis: OC-4 and Zurich Instruments: HF2-LI and HF2-PLL).~\cite{Alb91} For the STM measurement, the bias voltage was applied to the tip while the sample was electronically grounded. The tip apex was ex situ sharpened by milling with a focused ion beam. The tip radius was less than 10~nm. A clean gold tip was in situ formed by indenting to the Au sample surface and applying a pulse bias voltage between tip and sample several times. For AFM, the tip apex was terminated with a CO molecule, which was picked up from the surface.~\cite{Bar97} Clean Au(111) surfaces were in situ prepared by repeated cycles of standard Ar$^+$ sputtering ($3 \times 10^{-6}$ mbar, 1000 eV, and 15 min) and annealing at 720 K. 
7-AGNRs were grown under ultrahigh vacuum conditions (base pressure $1 \times 10^{-10} mbar$) using 10,10‘-dibromo-9,9‘-bianthryl (DBBA) as precursor monomer, as reported previously \cite{Cai2010}. After deposition of DBBA on the Au(111) surface held at room temperature, polymerization and subsequent cyclodehydrogenation were induced by annealing to 300 deg C or 400 deg C to obtain a majority of CH$_2$- or CH-terminated 7-AGNRs, respectively (see text for discussion of ribbon termination). Identification of termination was accomplished by scanning at low bias where CH-terminated 7-AGNRs exhibit a specific density of states pattern at the ribbon termini \cite{Talirz2013}.


\bibliography{library}

\end{document}